\documentclass[%
 reprint,
 amsmath,amssymb,
 aps,
 prd,
]{revtex4-2}

\usepackage{CJK}
\usepackage{graphicx}
\usepackage{dcolumn}
\usepackage{bm}
\usepackage{aas_macros}
\usepackage{color}

\usepackage{natbib}

\newcommand{\bmx}{\bm{x}}
\newcommand{\bmk}{\bm{k}}

\begin{document}


\title{Systematic analysis of Parity-Violating modes}

\author{Hong-Ming Zhu \begin{CJK*}{UTF8}{gbsn}(朱弘明)\end{CJK*}}
\email{hmzhu@nao.cas.cn}
\affiliation{National Astronomical Observatories, Chinese Academy of Sciences, 20A Datun Road, Beijing 100101, China}
\affiliation{Canadian Institute for Theoretical Astrophysics, 60 St. George Street, Toronto, Ontario M5S 3H8, Canada}

\author{Ue-Li Pen \begin{CJK*}{UTF8}{bsmi}(彭威禮)\end{CJK*}}
\email{pen@cita.utoronto.ca}
\affiliation{Institute of Astronomy and Astrophysics, Academia Sinica, Astronomy-Mathematics Building, No. 1, Sec. 4, Roosevelt Road, Taipei 10617, Taiwan}
\affiliation{Canadian Institute for Theoretical Astrophysics, 60 St. George Street, Toronto, Ontario M5S 3H8, Canada}
\affiliation{Dunlap Institute for Astronomy and Astrophysics, University of Toronto, 50 St. George Street, Toronto, Ontario M5S 3H4, Canada}
\affiliation{Canadian Institute for Advanced Research, 661 University Ave, Toronto, Ontario M5G 1M1, Canada}
\affiliation{Perimeter Institute for Theoretical Physics, 31 Caroline Street North, Waterloo, Ontario N2L 2Y5, Canada}

\date{\today}   

\begin{abstract}
Recent reports of cosmological parity violation in the 4PCF raises the
question of how such violations could be systematically generated.
Here we present a constructive procedure to generate arbitrary
violations of vectorial and tensorial types on any scale,
which is computationally efficient in the squeezed limit.
We directly
compute their numerical transfer function, and find strong
conservation in the linear regime.  This procedure spans all squeezed
parity violating observables at the 4PCF, following the quadratic estimator classification.
\end{abstract}


\maketitle


Testing parity violation involves examining the behavior of physical processes under spatial inversion.
In the realm of particle physics, parity violation was famously observed in weak interactions \cite{PhysRev.104.254,PhysRev.105.1413}, a discovery with profound implications that has transformed our understanding of fundamental symmetries in nature.
In cosmology, where macroscopic physics typically exhibit parity symmetry, any detected parity asymmetry provides a direct window into the inflationary Universe. 
Therefore, testing parity violation through cosmological observations is crucial for probing the fundamental symmetries of the physical laws governing the early Universe 
(e.g., \cite[][]{1999PhRvL..83.1506L,2008PhRvL.101n1101C,2009PhRvL.102w1301T,2011JCAP...06..003S,2020JHEP...04..189L,2023JHEP...02..021C}, and reference therein).

Recent reports of evidence and upper bound for parity violation in large-scale structure (LSS) and cosmic microwave background (CMB) observations \cite[][]{2023MNRAS.522.5701H,2022PhRvD.106f3501P,2024arXiv240109523P,2024arXiv240504660A,2024arXiv240703397K,2023PhRvL.131r1001P,2024PhRvD.109h3514P,2022PhRvD.105h3512M,2020PhRvL.125v1301M,2022PhRvL.128i1302D,2022PhRvD.106f3503E} have motivated the question of how such parity violations could be generated.
A broad class of parity violations is due to gravitational fossils
\cite{2010PhRvL.105p1302M,2012PhRvL.108y1301J,2017PhRvL.118v1301M}.  
These can be tensorial, for example arising from decaying primordial gravitational waves, or vectorial, as expected from axionic mechanisms.

One approach to generating these is through explicit parity-violating dynamics (e.g., Quijote \cite{2024PhRvD.109b3531C}, etc.).  
Such procedures generate substantial perturbations even at the level of the two-point correlation function (2PCF) and are fully non-Gaussian in the one-point correlation function (1PCF),  exhibiting non-zero kurtosis.
In contrast, gravitational fossils are initial conditions that remain explicitly Gaussian in the 1PCF, whose parity violation are akin to CMB lensing. 
Vectorial perturbations can be constructed using a lensing remap procedure \cite{2024arXiv240606080S}, but a coordinate change in Euclidean space is at most vector in nature.

In this paper, we present a new method to generate parity-violating initial conditions with arbitrary vectorial and tensorial types of violations. 
Utilizing $N$-body simulations, we investigate the impact of gravitational nonlinear evolution on the observability of a parity-violating primordial Universe.
We find that tidal estimators can efficiently distinguish the parity-violating signal from gravitational nonlinearities generated in the late Universe, because of the different symmetry properties exhibited under parity transformation.

These findings open up new possibilities for exploring the Universe's fundamental symmetries and have profound implications for detecting parity-violating physics using galaxy distributions.


{\it Helical vector and tensor modes.}---We consider vector and tensor perturbations in the linearized spatial metric $h_{ij}(\bmx)$.
These perturbations can be sourced by inflationary dynamics, where quantum fluctuations in the early Universe give rise to primordial gravitational waves (tensor modes) and potentially vector modes, depending on the specific model of inflation.
In Fourier space, $h_{ij}(\bm{k})=\int \mathrm{d}^3\bm{x} e^{-i\bm{k}\cdot\bmx}h_{ij}(\bmx)$.

The vector modes can be described by a divergence-free, transverse vector field
$v_a(\bmx)$, which has two degrees of freedom.
In Fourier space, these two degrees of freedom can be described by the circular polarization amplitudes $v_R$ and $v_L$ as
\begin{equation}
    v_{i}(\bmk)=v_R(\bmk)\, e^R_{i}(\hat{\bmk}) + v_L(\bmk)\, e^L_{i}(\hat{\bmk}),
\end{equation}
where $e^\lambda_i(\hat{\bmk})$ are unit polarization vectors.
Vector perturbations in the metric $h_{ij}(\bm{k})$ is given by the derivative of the vector field, $i\hat{k}_i v_j + i\hat{k}_j v_i$.
The vector power spectrum is defined by \(\langle v_i(\bm{k}) v_i^*(\bm{k}') \rangle = (2\pi)^3 P_v(k) \delta^{(3)}(\bm{k} - \bm{k}')
\).

Gravitational waves or tensor modes are encoded in the transverse, traceless part of the spatial metric, which has two degrees of freedom.
In Fourier space, these two degrees of freedom can be described by the circular polarization amplitudes $h_R$ and $h_L$ as
\begin{equation}
    h_{ij}(\bm{k})=h_R(\bm{k}) e^R_{ij}(\hat{\bm{k}}) + h_L(\bm{k}) e^L_{ij}(\hat{\bm{k}}),
\end{equation}
where the polarization tensors \(e^{\lambda}_{ab}(\hat{\bmk})\) are normalized so that \(e^{\lambda}_{ij}(\hat{\bmk})e^{\lambda'}_{ij}(\hat{\bmk})=2\delta_{\lambda\lambda'}\).
The tensor power spectrum is defined by \(\langle h_{ij}(\bm{k}) h_{ij}^*(\bm{k}') \rangle = (2\pi)^3 P_h(k) \delta^{(3)}(\bm{k} - \bm{k}')\).

Large-scale vector and tensor perturbations leave a local anisotropic tidal imprint in the smaller-scale matter distribution, which persists long after the large-scale modes have decayed, as introduced by Masui and Pen \cite{2010PhRvL.105p1302M}.
Therefore, this imprint constitutes a fossilized map of the vector and tensor modes, providing a unique window into the new physics of the early Universe. 
More comprehensive treatments of the fossil effect were provided by Refs~\cite{2013PhRvD..88d3507D,2014PhRvD..89h3507S,2023PhRvD.107f3531A}.

The effect of large-scale metric perturbations on the locally measured small-scale matter power spectrum is given by 
\begin{equation}
\label{eq:pk_hij}
    P_\delta(\bmk,\tau)|_{h_{ij}}=P_\delta(k,\tau)+\hat{k}^i\hat{k}^jh_{ij}^{(0)}f(k,\tau)P_\delta(k,\tau),
\end{equation}
with
\begin{equation}
\label{eq:response}
    f(k,\tau)=2\alpha(\tau)-\beta(\tau)\frac{\mathrm{d}\ln P_\delta(k,\tau)}{\mathrm{d}\ln k},
\end{equation}
where  $P(k,\tau)$ represents the linear matter power spectrum evaluated at conformal time $\tau$, and $h_{ab}^{(0)}$ is the primordial amplitude of metric perturbations.
The function $\alpha(\tau)$ describes the growth of tidal interaction and is cosmology dependent \cite{2014PhRvD..89h3507S}, while $\beta(\tau)$ quantifies the effect of local anisotropic dilation on the coordinates and asymptotes to $1/2$ when the tensor or vector mode has long entered the horizon and decayed.

Such effects can be used to search for parity violation was first suggested by Jeong and Kamionkowski \cite{2012PhRvL.108y1301J}.
Masui {\it et al.} \cite{2017PhRvL.118v1301M} presented a detailed study of probing parity violation in primordial gravitational waves through the gravitational fossil effect.
A variety of sources of gravitational parity violation could have left an imprint on the net helicity of vector and tensor perturbations \cite[e.g.,][]{2018PhRvD..97b3532C}, namely through the preferred excitation of one circular polarization over the other.
Such a preferred handedness leaves a tidal imprint on the large-scale structure, preserving the information about the primordial vector and tensor helicities.

However, the above expression is valid only for $k$ in the linear regime.
Nonlinear gravitational evolution generates complex effects on small scales that can obscure the fossilized imprints left by primordial vector and tensor perturbations.
This is essentially why numerical simulations are required to assess the impact of nonlinear gravitational effects on the detectability of the fossil effect, as we will now explicitly demonstrate.

{\it Helical parity-violating initial conditions.}---Vector and tensor modes are perturbations to the spatial metric.
An arbitrary metric can contain intrinsic curvature, making it irreducible to Euclidean space.
To account for the effects helical vector and tensor modes, it is necessary to introduce a perturbed spatial metric $h_{ab}$, which encompasses these perturbations.

A general algorithm for implementing a non-stationary Gaussian field involves using the metric $h_{ab}$ 
to calculate the geodesic distance between pairs of points, incorporating the effects of metric perturbations into the density covariance matrix, and then diagonalizing this matrix \cite{2020arXiv201108251Z}.
Gaussian random numbers are then generated with variances corresponding to the eigenvalues to create a realization of non-stationary Gaussian field.

This method allows us to create arbitrary violations of vectorial and tensorial types on any scale.
However, this diagonalization cannot be performed using a fast Fourier transform, significantly increasing the computational cost.
In the squeezed limit, where the wavelength of vector and tensor modes are much larger than the small-scale density modes, the metric perturbations $h_{ab}$ with wavenumber $k$ can be considered spatially constant in a region of size $R\ll1/k$.
This simplification allows for a computational shortcut by using fast Fourier transform to compute convolutions \cite{1997ApJ...490L.127P,2006NewA...11..273T}, without the need for full-scale diagonalization.
In this paper, we adopt the latter efficient approach and plan to investigate the general algorithm in future work.

In the standard approach, the initial density field $\delta(\bmx)$ is generated by convolving a realization of Gaussian white noise fluctuations $n(\bmx)$ with a kernel function $w(\bmx)$,
\begin{equation}
    \delta(\bmx)=\int d^3x' n(\bmx')w(\bmx-\bmx'),
\end{equation}
where $w(r)$ is the Fourier transform of the square root of the linear power spectrum $P_\delta(k)$.

Using the two-level mesh scheme from Refs.~\cite{1997ApJ...490L.127P,2006NewA...11..273T}, we decompose the kernel $w(r)$ into a short-range component $w_s(r)$ and a long-range component $w_l(r)$ as
\begin{equation}
    w(r)=w_s(r)+w_l(r),   
\end{equation}
where
\begin{align}
    w_s(r) &= \begin{cases}
        w(r) - \alpha(r) & \text{if } r \leq r_d, \\
        0 & \text{if } r > r_d.
    \end{cases}
\end{align}
and
\begin{align}
    w_l(r) &= \begin{cases}
        \alpha(r) & \text{if } r \leq r_d, \\
        w(r) & \text{if } r > r_d.
    \end{cases}
\end{align}
Here $r_d$ is a free parameter that defines the short range cut off.
The function \(\alpha(r)\) is chosen to be
\begin{equation}
    \alpha(r) = (a r^2 + b r^4 + c r^6) w(r),
\end{equation}
where coefficients $a = 3/r_d^2$, $b = -3/r_d^4$, and $c = 1/r_d^6$ are determined from the conditions $\alpha(r_d) = w(r_d)$, $\alpha'(r_d) = w'(r_d)$, and $\alpha''(r_d) = w''(r_d)$ \cite{2006NewA...11..273T}.

The convolution of the short-range kernel can be performed locally within a cubic subvolume, including a buffer zone of thickness greater than $r_d$ \cite{1997ApJ...490L.127P,2006NewA...11..273T}.
Long-wavelength metric perturbations $h_{ij}$ can be treated as spatially constant when the wavelength is much larger the subvolume size.
To incorporate the fossil effect in the initial conditions, we can use the anisotropic kernel
\begin{equation}
    \widetilde{w}_s(\bm{r})=w_s\left(\sqrt{g_{ij}r^ir^j}\right),
\end{equation}
where the metric $g_{ij}=\delta_{ij}+h_{ij}$ takes the local constant value in this subvolume.
This form accounts for the anisotropic nature of the fossil effect by modifying the distance measure in the kernel through the metric $g_{ij}$.
The local small-scale power spectrum is given by the square of the Fourier transform of \(\widetilde{w}_s(\bm{r})\),
expressed as
\begin{equation}
    \widetilde{w}_s^2(\bmk) = w_s^2(k) - \frac{1}{2} \hat{k}^i\hat{k}^jh_{ij}w_s^2(k)\frac{\mathrm{d}\ln w_s^2(k)}{\mathrm{d}\ln k} + \cdots,
\end{equation}
where the second term directly corresponds to the one proportional to $\beta(\tau)$ in Eq.~(\ref{eq:pk_hij}).

Notice that \(\mathrm{d}\ln P_\delta(k)/\mathrm{d}\ln k\) is nearly constant on small scales, where the power spectrum follows a power law.
In this regime, both terms in Eq.~(\ref{eq:response}) are scale-independent and have the same effect on the local anisotropic small-scale power spectrum.
The response of small-scale density perturbations to large-scale metric perturbations is scale-invariant in the squeezed limit.
The term involving \(\alpha(\tau)\) can be easily accounted for by adjusting the amplitude of \(h_{ij}\) in our method.
However, we do not consider this term in this paper, as the \(\beta(\tau)\) term already captures the primary effect.

In the numerical implementation, we generate density perturbations on a \(512^3\) grid within a cubic box of size \(L = 800\,\mathrm{Mpc}/h\), and long-wavelength metric perturbations on a \(64^3\) grid of subvolumes, which we refer to as tiles, within the same volume.
Each tile is extended by a buffer region with a thickness of 10 grid cells. 
This buffer zone ensures that all necessary contributions to the convolution are captured within the subvolume, accounting for the kernel's spatial extent, 
where \(r_d\) corresponds to 8 grid cells \cite{2013MNRAS.436..540H}.
The short-range kernels are calculated with no boundary effects imposed by the domain decomposition.
We generate the helical vector and tensor perturbations using non-scale-invariant power spectra, with \(P_v(k) = A_v\) and \(P_h(k) = A_h\) for \(|k - k_0| < k_\mathrm{f}/2\), and zero otherwise, where \(k_\mathrm{f} = 2\pi/L\) is the fundamental frequency.
In this setup, we take \(k_0 = 4k_\mathrm{f}\), corresponding to a wavelength of \(L/4\), which is 16 times larger than the characteristic size of a tile, \(L/64\).
We assume maximal helicity, characterized by a helicity asymmetry of \(\Delta\chi \equiv (P_{R} - P_{L}) / (P_{R} + P_{L}) = 1\), indicating a fully right-handed polarization.
The power spectrum amplitude is set so that the variance \(\langle h_{ij}(\bm{x}) h^{ij}(\bm{x}) \rangle / 8 = 0.01\), corresponding to a \(10\%\) local anisotropic shear.
We start from initial redshift $z_\mathrm{init}=99$ and evolve $512^3$ dark matter particles to redshift $z=0$.
The density field is generated using the cloud-in-cell interpolation scheme.

\begin{figure*}[htb!]
\centering
\includegraphics[width=0.32\textwidth]{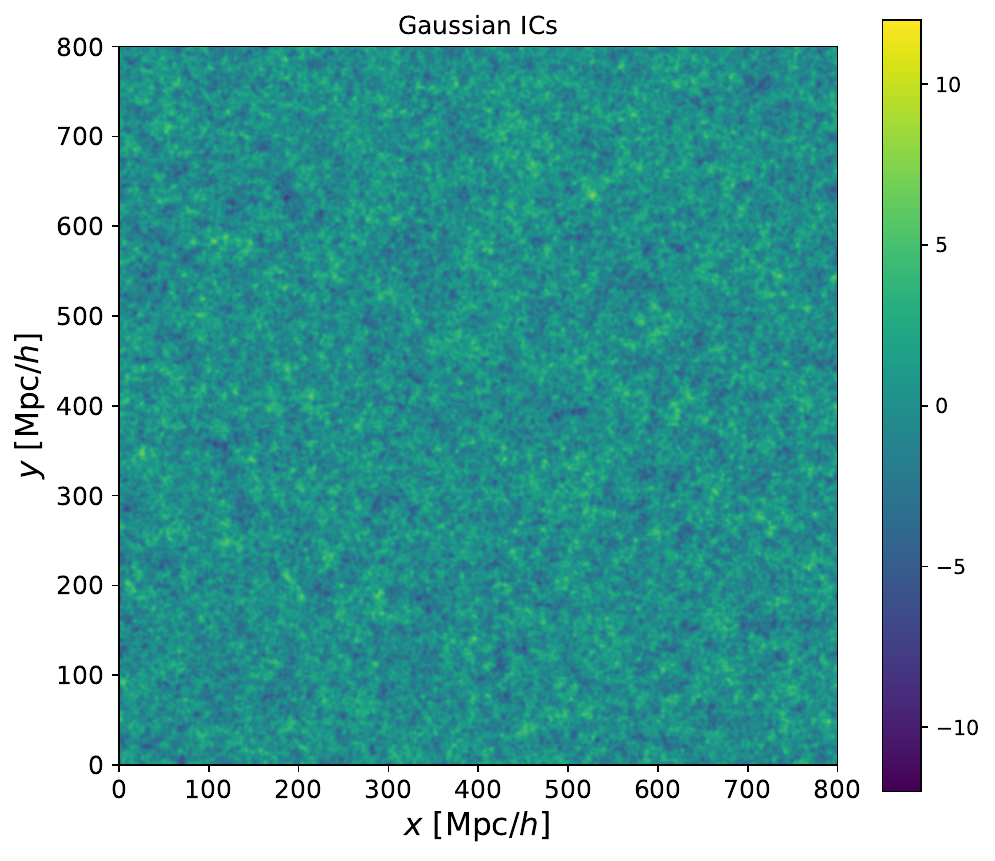}
\includegraphics[width=0.32\textwidth]{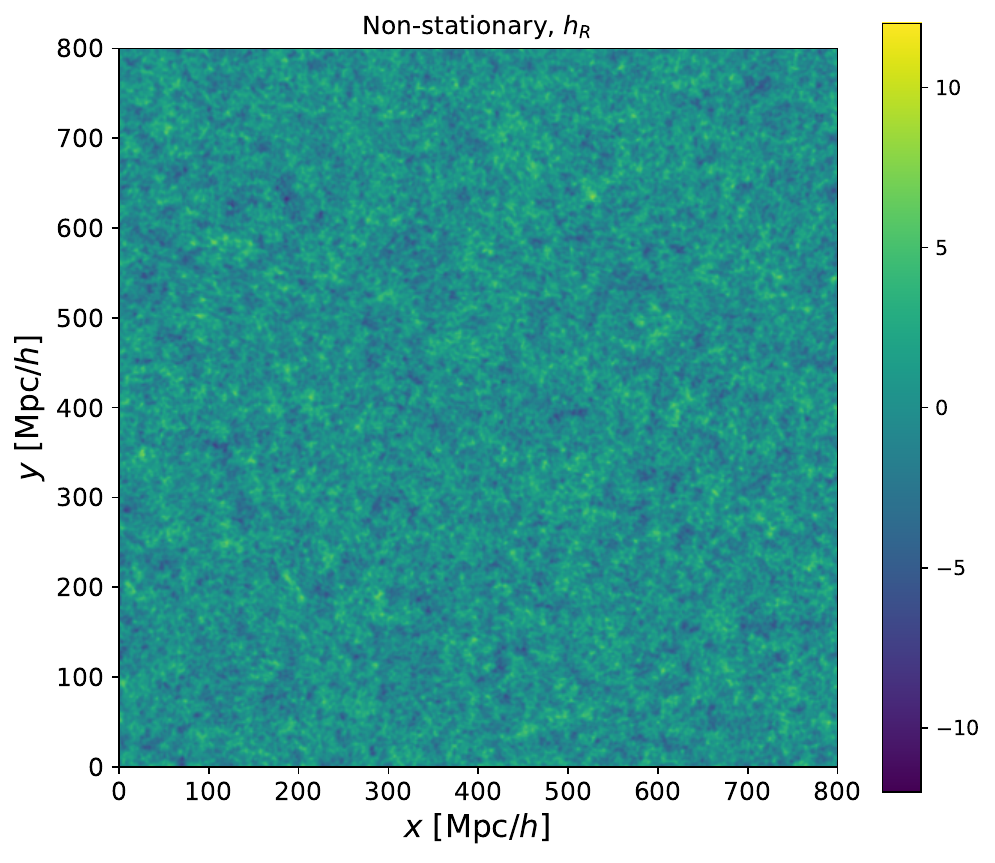}
\includegraphics[width=0.32\textwidth]{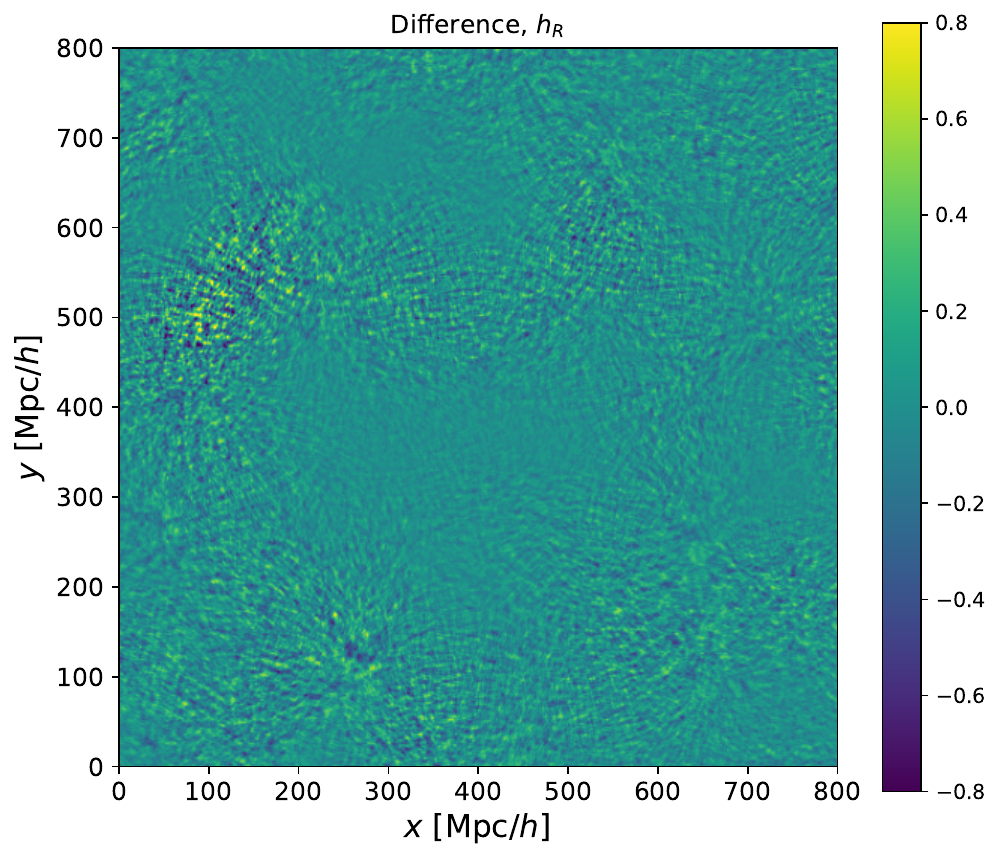}
\caption{Two-dimensional maps of Gaussian initial conditions (left), non-stationary initial conditions with helical tensor modes (middle), and their differences (right), each with a thickness of $1.56\,h^{-1}\mathrm{Mpc}$.
The density fields are smoothed using a Gaussian window with a scale $1\,h^{-1}\mathrm{Mpc}$.
The differences clearly reveal the presence of curl patterns, highlighting the distinction between the helical and Gaussian initial conditions.\label{fig:map}}
\end{figure*}

In Fig.~\ref{fig:map},
we present two-dimensional maps of Gaussian initial conditions, non-stationary initial conditions with helical tensor modes, and their differences, each with a thickness of $1.56\,h^{-1}\mathrm{Mpc}$.
The density fields are smoothed using a Gaussian window with a scale $1\,h^{-1}\mathrm{Mpc}$.
The differential map clearly reveals the presence of curl patterns, highlighting the distinction between the helical and Gaussian initial conditions.

Our method generates non-stationary initial conditions, while having minimal impact on the power spectrum.
Even with a $10\%$ local shear, the effect on power spectrum is limited to only about $1\%$.
The initial conditions remain Gaussian, as the local anisotropic convolution, being a linear operation, does not alter the statistical Gaussianity, ensuring that the 1PCF stays fully Gaussian even in the presence of local anisotropy.

{\it Tidal estimators.}---Local anisotropic distortions in the matter distribution have been extensively investigated to reconstruct the large-scale longitudinal scalar mode from small-scale density fluctuations \cite{2012arXiv1202.5804P,2016PhRvD..93j3504Z,2019MNRAS.486.3864K,2022ApJ...929....5Z,2024ApJ...962...21Z} (see also Refs.~\cite{2018JCAP...07..046F,2020PhRvD.101h3510L,2020arXiv200700226L,2021PhRvD.104l3520D,2024JCAP...07..020W}).
In three dimension, the tidal shear tensor has five degrees of freedom. 
Beyond the large-scale scalar tidal field, the remaining four degrees of freedom correspond to two vector modes and two tensor modes.

To extract information from the vector and tensor fossil effects, optimal estimators for these modes can be constructed, as detailed by Jeong and Kamionkowski \cite{2012PhRvL.108y1301J}.
The optimal estimator for vector modes is
\begin{equation}
\hat{v}_\lambda(\bm{K}) = N^V(K) \sum_{\bm{k}} \frac{-i\hat{K}^ae_{b}^{\lambda*}(\hat{\bm{K}}) \hat{k}^a \hat{k}^b f(k)P(k)}{2V P^{\text{tot}}(k) P^{\text{tot}}(|\bm{K} - \bm{k}|)} \delta(\bm{k}) \delta(\bm{K} - \bm{k}),
\label{eq:helical_vector_estimator}
\end{equation}
and the optimal estimator for tensor modes is given by 
\begin{equation}
\hat{h}_\lambda(\bm{K}) = N^T(K) \sum_{\bm{k}} \frac{e_{ab}^{\lambda*}(\hat{\bm{K}}) \hat{k}^a \hat{k}^b f(k)P(k)}{2V P^{\text{tot}}(k) P^{\text{tot}}(|\bm{K} - \bm{k}|)} \delta(\bm{k}) \delta(\bm{K} - \bm{k}).
\label{eq:helical_tensor_estimator}
\end{equation}
The above estimators can be constructed to be unbiased for nearly Gaussian fields by computing the noise power spectra $N^V$ and $N^T$, following the same procedure used in CMB lensing. 
However, in more general cases, such as nonlinear density fields, there is no analytical form for $N^V$ or $N^T$, and the above estimators cannot be normalized in the same way as for CMB lensing.
Therefore, we need to use simulations to calibrate these estimators by computing the cross-correlation with the input vector or tensor fields and scaling the estimated fields for normalization.

\begin{figure}[htb!]
\centering
\includegraphics[width=\columnwidth]{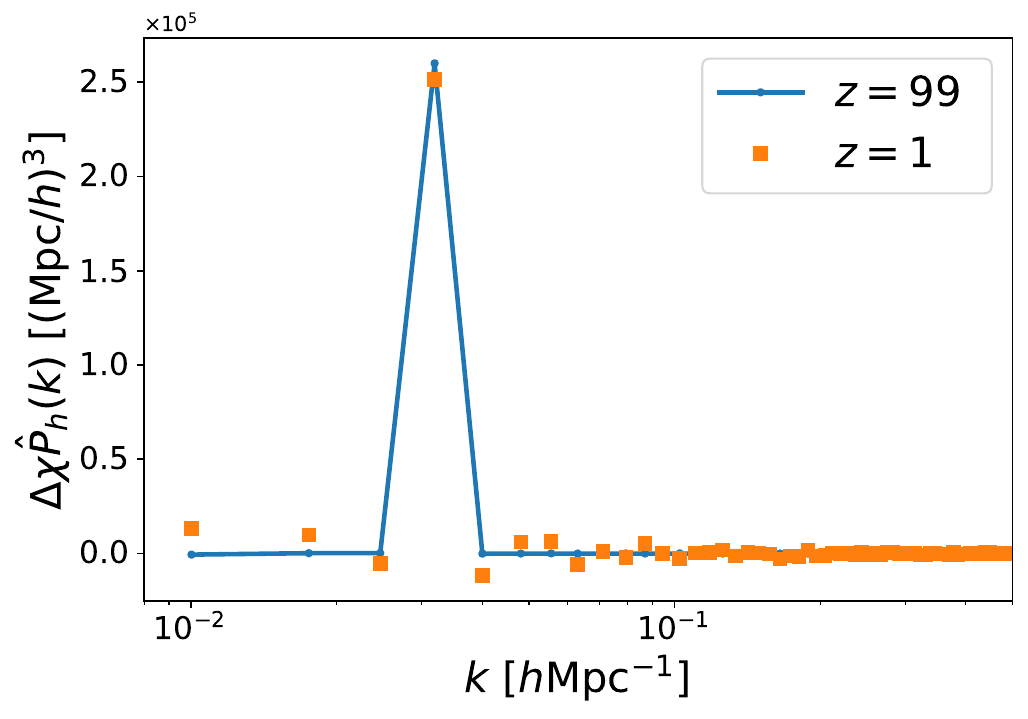}
\caption{\label{fig:tensor} Helicity power spectrum of tensor modes for $z=99$ and $z=1$.
We find strong conservation in the linear regime.
}
\end{figure}

\begin{figure}[htb!]
\centering
\includegraphics[width=\columnwidth]{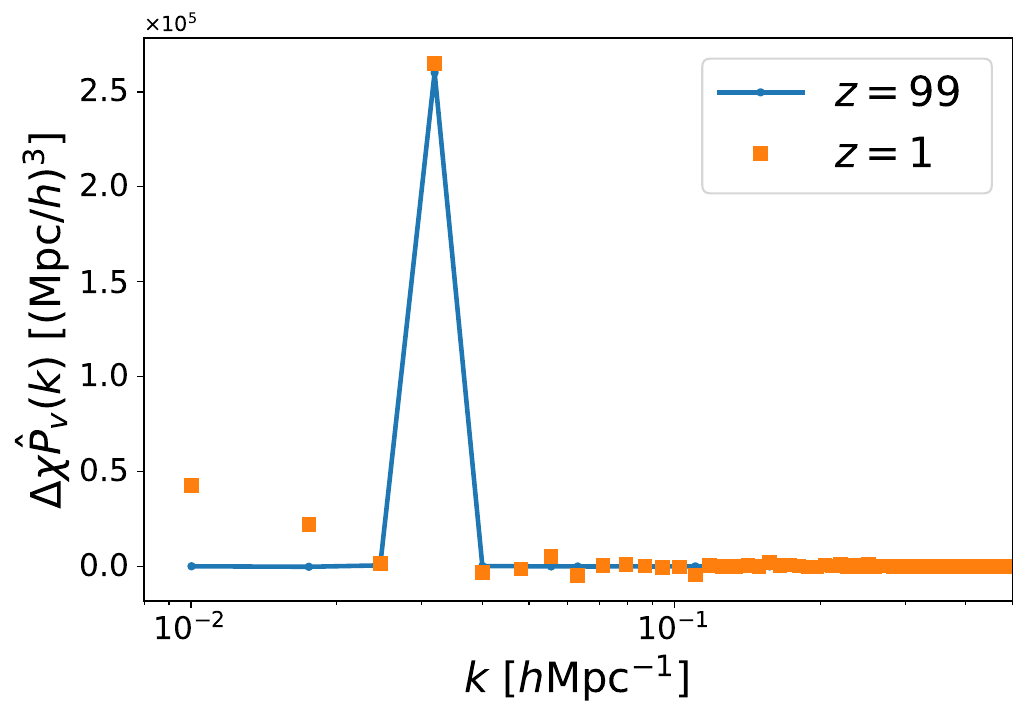}
\caption{\label{fig:vector} Same as Fig.~\ref{fig:tensor}, but for vector modes.}
\end{figure}

Figure~\ref{fig:tensor}
shows the helicity power spectrum for tensor modes $\Delta\chi \hat{P}_h(k)$ at $z=99$ and $z=1$.
At high redshift, $z=99$, the helicity signal is clearly visible, indicating the presence of a strong helical gravitational wave background around$k=4k_\mathrm{f}$. 
At lower redshift, the nonlinear evolution generates additional power and introduces random fluctuations into the helicity power spectrum. 
However, the nonlinear evolution does not generate a systematic helicity between the left- and right-handed modes, as the gravity in the simulations preserves parity symmetry. 
Moreover, the primordial helicity signal remains clearly visible at $z=1$, demonstrating that helicity is a preserved primordial feature, not created by late-time nonlinearities.
Figure~\ref{fig:vector}
shows the corresponding results for vector modes
$\Delta\chi \hat{P}_v(k)$, leading to similar conclusion.

{\it Conclusions.}---We present a new method to generate parity-violating initial conditions of vector and tensor types for cosmological simulations.
This approach spans all
parity-violating observables at the level of 4PCF in the squeezed limit, following the quadratic estimator classification introduced by Jeong and Kamionkowski \cite{2012PhRvL.108y1301J}.

Akitsu {\it et al.} \cite{2023PhRvD.107f3531A} have included the gravitational wave in the simulation background, however, the global background does not carry helicity information. 

We use $N$-body simulations to investigate the impact of nonlinear evolution on the detectability of the primordial helicity and numerically measure the transfer function of the helicity power spectrum.
While nonlinear evolution introduces extra stochasticities,
tidal estimators can efficiently separate parity-odd signals from parity-even nonlinearities generated in the late Universe, due to the different symmetry properties exhibited by physical systems under parity transformation.

These findings open new avenues for probing parity-violating new physics using galaxy surveys.
The method enables us to quantify the response of estimators to primordial helical vector and tensor modes. Moreover, it allows us to direct constrain the amplitude of the vector and tensor helicity power spectrum, rather than relying solely on statistical tests.

This method is not only limited to testing the sensitivity of tidal estimators, but can also be applied to a range of other approaches, including 4PCF \cite{Cahn:2021ltp}, galaxy spin \cite{2020PhRvL.124j1302Y}, galaxy shape \cite{2024arXiv240504210O}, parity-odd power spectra \cite{2024arXiv240615683J}, and machine learning techniques \cite{2024PhRvD.109h3518T,2024arXiv240513083C}, which is of interest to the broader community.

{\it Acknowledgement.---}HMZ thanks the organizers and participants of the ``Cosmology in the Adriatic'' workshop for many enlightening discussions that contributed to this work. 
HMZ receives support from the National Key R\&D Program of China (2023YFA1607800).
ULP receives support from the Canadian Institute for Advanced Research (CIFAR) Fund, the Natural Sciences and 
Engineering Research Council of Canada (NSERC) [funding 
reference numbers RGPIN-2019-06770, ALLRP 586559-23], 
and the Ontario Research Fund-Research Excellence (ORF-RE) Fund.
Research at the Perimeter Institute is supported by the Government of Canada 
through Industry Canada and by the Province of Ontario through the Ministry of 
Research and Innovation.


\bibliography{apssamp}

\providecommand{\noopsort}[1]{}\providecommand{\singleletter}[1]{#1}%
\begin{thebibliography}{48}%
\makeatletter
\providecommand \@ifxundefined [1]{%
 \@ifx{#1\undefined}
}%
\providecommand \@ifnum [1]{%
 \ifnum #1\expandafter \@firstoftwo
 \else \expandafter \@secondoftwo
 \fi
}%
\providecommand \@ifx [1]{%
 \ifx #1\expandafter \@firstoftwo
 \else \expandafter \@secondoftwo
 \fi
}%
\providecommand \natexlab [1]{#1}%
\providecommand \enquote  [1]{``#1''}%
\providecommand \bibnamefont  [1]{#1}%
\providecommand \bibfnamefont [1]{#1}%
\providecommand \citenamefont [1]{#1}%
\providecommand \href@noop [0]{\@secondoftwo}%
\providecommand \href [0]{\begingroup \@sanitize@url \@href}%
\providecommand \@href[1]{\@@startlink{#1}\@@href}%
\providecommand \@@href[1]{\endgroup#1\@@endlink}%
\providecommand \@sanitize@url [0]{\catcode `\\12\catcode `\$12\catcode `\&12\catcode `\#12\catcode `\^12\catcode `\_12\catcode `\%12\relax}%
\providecommand \@@startlink[1]{}%
\providecommand \@@endlink[0]{}%
\providecommand \url  [0]{\begingroup\@sanitize@url \@url }%
\providecommand \@url [1]{\endgroup\@href {#1}{\urlprefix }}%
\providecommand \urlprefix  [0]{URL }%
\providecommand \Eprint [0]{\href }%
\providecommand \doibase [0]{https://doi.org/}%
\providecommand \selectlanguage [0]{\@gobble}%
\providecommand \bibinfo  [0]{\@secondoftwo}%
\providecommand \bibfield  [0]{\@secondoftwo}%
\providecommand \translation [1]{[#1]}%
\providecommand \BibitemOpen [0]{}%
\providecommand \bibitemStop [0]{}%
\providecommand \bibitemNoStop [0]{.\EOS\space}%
\providecommand \EOS [0]{\spacefactor3000\relax}%
\providecommand \BibitemShut  [1]{\csname bibitem#1\endcsname}%
\let\auto@bib@innerbib\@empty
\bibitem [{\citenamefont {Lee}\ and\ \citenamefont {Yang}(1956)}]{PhysRev.104.254}%
  \BibitemOpen
  \bibfield  {author} {\bibinfo {author} {\bibfnamefont {T.~D.}\ \bibnamefont {Lee}}\ and\ \bibinfo {author} {\bibfnamefont {C.~N.}\ \bibnamefont {Yang}},\ }\href {https://doi.org/10.1103/PhysRev.104.254} {\bibfield  {journal} {\bibinfo  {journal} {Phys. Rev.}\ }\textbf {\bibinfo {volume} {104}},\ \bibinfo {pages} {254} (\bibinfo {year} {1956})}\BibitemShut {NoStop}%
\bibitem [{\citenamefont {Wu}\ \emph {et~al.}(1957)\citenamefont {Wu}, \citenamefont {Ambler}, \citenamefont {Hayward}, \citenamefont {Hoppes},\ and\ \citenamefont {Hudson}}]{PhysRev.105.1413}%
  \BibitemOpen
  \bibfield  {author} {\bibinfo {author} {\bibfnamefont {C.~S.}\ \bibnamefont {Wu}}, \bibinfo {author} {\bibfnamefont {E.}~\bibnamefont {Ambler}}, \bibinfo {author} {\bibfnamefont {R.~W.}\ \bibnamefont {Hayward}}, \bibinfo {author} {\bibfnamefont {D.~D.}\ \bibnamefont {Hoppes}},\ and\ \bibinfo {author} {\bibfnamefont {R.~P.}\ \bibnamefont {Hudson}},\ }\href {https://doi.org/10.1103/PhysRev.105.1413} {\bibfield  {journal} {\bibinfo  {journal} {Phys. Rev.}\ }\textbf {\bibinfo {volume} {105}},\ \bibinfo {pages} {1413} (\bibinfo {year} {1957})}\BibitemShut {NoStop}%
\bibitem [{\citenamefont {{Lue}}\ \emph {et~al.}(1999)\citenamefont {{Lue}}, \citenamefont {{Wang}},\ and\ \citenamefont {{Kamionkowski}}}]{1999PhRvL..83.1506L}%
  \BibitemOpen
  \bibfield  {author} {\bibinfo {author} {\bibfnamefont {A.}~\bibnamefont {{Lue}}}, \bibinfo {author} {\bibfnamefont {L.}~\bibnamefont {{Wang}}},\ and\ \bibinfo {author} {\bibfnamefont {M.}~\bibnamefont {{Kamionkowski}}},\ }\href {https://doi.org/10.1103/PhysRevLett.83.1506} {\bibfield  {journal} {\bibinfo  {journal} {\prl}\ }\textbf {\bibinfo {volume} {83}},\ \bibinfo {pages} {1506} (\bibinfo {year} {1999})},\ \Eprint {https://arxiv.org/abs/astro-ph/9812088} {arXiv:astro-ph/9812088 [astro-ph]} \BibitemShut {NoStop}%
\bibitem [{\citenamefont {{Contaldi}}\ \emph {et~al.}(2008)\citenamefont {{Contaldi}}, \citenamefont {{Magueijo}},\ and\ \citenamefont {{Smolin}}}]{2008PhRvL.101n1101C}%
  \BibitemOpen
  \bibfield  {author} {\bibinfo {author} {\bibfnamefont {C.~R.}\ \bibnamefont {{Contaldi}}}, \bibinfo {author} {\bibfnamefont {J.}~\bibnamefont {{Magueijo}}},\ and\ \bibinfo {author} {\bibfnamefont {L.}~\bibnamefont {{Smolin}}},\ }\href {https://doi.org/10.1103/PhysRevLett.101.141101} {\bibfield  {journal} {\bibinfo  {journal} {\prl}\ }\textbf {\bibinfo {volume} {101}},\ \bibinfo {eid} {141101} (\bibinfo {year} {2008})},\ \Eprint {https://arxiv.org/abs/0806.3082} {arXiv:0806.3082 [astro-ph]} \BibitemShut {NoStop}%
\bibitem [{\citenamefont {{Takahashi}}\ and\ \citenamefont {{Soda}}(2009)}]{2009PhRvL.102w1301T}%
  \BibitemOpen
  \bibfield  {author} {\bibinfo {author} {\bibfnamefont {T.}~\bibnamefont {{Takahashi}}}\ and\ \bibinfo {author} {\bibfnamefont {J.}~\bibnamefont {{Soda}}},\ }\href {https://doi.org/10.1103/PhysRevLett.102.231301} {\bibfield  {journal} {\bibinfo  {journal} {\prl}\ }\textbf {\bibinfo {volume} {102}},\ \bibinfo {eid} {231301} (\bibinfo {year} {2009})},\ \Eprint {https://arxiv.org/abs/0904.0554} {arXiv:0904.0554 [hep-th]} \BibitemShut {NoStop}%
\bibitem [{\citenamefont {{Sorbo}}(2011)}]{2011JCAP...06..003S}%
  \BibitemOpen
  \bibfield  {author} {\bibinfo {author} {\bibfnamefont {L.}~\bibnamefont {{Sorbo}}},\ }\href {https://doi.org/10.1088/1475-7516/2011/06/003} {\bibfield  {journal} {\bibinfo  {journal} {\jcap}\ }\textbf {\bibinfo {volume} {2011}},\ \bibinfo {eid} {003} (\bibinfo {year} {2011})},\ \Eprint {https://arxiv.org/abs/1101.1525} {arXiv:1101.1525 [astro-ph.CO]} \BibitemShut {NoStop}%
\bibitem [{\citenamefont {{Liu}}\ \emph {et~al.}(2020)\citenamefont {{Liu}}, \citenamefont {{Tong}}, \citenamefont {{Wang}},\ and\ \citenamefont {{Xianyu}}}]{2020JHEP...04..189L}%
  \BibitemOpen
  \bibfield  {author} {\bibinfo {author} {\bibfnamefont {T.}~\bibnamefont {{Liu}}}, \bibinfo {author} {\bibfnamefont {X.}~\bibnamefont {{Tong}}}, \bibinfo {author} {\bibfnamefont {Y.}~\bibnamefont {{Wang}}},\ and\ \bibinfo {author} {\bibfnamefont {Z.-Z.}\ \bibnamefont {{Xianyu}}},\ }\href {https://doi.org/10.1007/JHEP04(2020)189} {\bibfield  {journal} {\bibinfo  {journal} {Journal of High Energy Physics}\ }\textbf {\bibinfo {volume} {2020}},\ \bibinfo {eid} {189} (\bibinfo {year} {2020})},\ \Eprint {https://arxiv.org/abs/1909.01819} {arXiv:1909.01819 [hep-ph]} \BibitemShut {NoStop}%
\bibitem [{\citenamefont {{Cabass}}\ \emph {et~al.}(2023)\citenamefont {{Cabass}}, \citenamefont {{Jazayeri}}, \citenamefont {{Pajer}},\ and\ \citenamefont {{Stefanyszyn}}}]{2023JHEP...02..021C}%
  \BibitemOpen
  \bibfield  {author} {\bibinfo {author} {\bibfnamefont {G.}~\bibnamefont {{Cabass}}}, \bibinfo {author} {\bibfnamefont {S.}~\bibnamefont {{Jazayeri}}}, \bibinfo {author} {\bibfnamefont {E.}~\bibnamefont {{Pajer}}},\ and\ \bibinfo {author} {\bibfnamefont {D.}~\bibnamefont {{Stefanyszyn}}},\ }\href {https://doi.org/10.1007/JHEP02(2023)021} {\bibfield  {journal} {\bibinfo  {journal} {Journal of High Energy Physics}\ }\textbf {\bibinfo {volume} {2023}},\ \bibinfo {eid} {21} (\bibinfo {year} {2023})},\ \Eprint {https://arxiv.org/abs/2210.02907} {arXiv:2210.02907 [hep-th]} \BibitemShut {NoStop}%
\bibitem [{\citenamefont {{Hou}}\ \emph {et~al.}(2023)\citenamefont {{Hou}}, \citenamefont {{Slepian}},\ and\ \citenamefont {{Cahn}}}]{2023MNRAS.522.5701H}%
  \BibitemOpen
  \bibfield  {author} {\bibinfo {author} {\bibfnamefont {J.}~\bibnamefont {{Hou}}}, \bibinfo {author} {\bibfnamefont {Z.}~\bibnamefont {{Slepian}}},\ and\ \bibinfo {author} {\bibfnamefont {R.~N.}\ \bibnamefont {{Cahn}}},\ }\href {https://doi.org/10.1093/mnras/stad1062} {\bibfield  {journal} {\bibinfo  {journal} {\mnras}\ }\textbf {\bibinfo {volume} {522}},\ \bibinfo {pages} {5701} (\bibinfo {year} {2023})},\ \Eprint {https://arxiv.org/abs/2206.03625} {arXiv:2206.03625 [astro-ph.CO]} \BibitemShut {NoStop}%
\bibitem [{\citenamefont {{Philcox}}(2022)}]{2022PhRvD.106f3501P}%
  \BibitemOpen
  \bibfield  {author} {\bibinfo {author} {\bibfnamefont {O.~H.~E.}\ \bibnamefont {{Philcox}}},\ }\href {https://doi.org/10.1103/PhysRevD.106.063501} {\bibfield  {journal} {\bibinfo  {journal} {\prd}\ }\textbf {\bibinfo {volume} {106}},\ \bibinfo {eid} {063501} (\bibinfo {year} {2022})},\ \Eprint {https://arxiv.org/abs/2206.04227} {arXiv:2206.04227 [astro-ph.CO]} \BibitemShut {NoStop}%
\bibitem [{\citenamefont {{Philcox}}\ and\ \citenamefont {{Ereza}}(2024)}]{2024arXiv240109523P}%
  \BibitemOpen
  \bibfield  {author} {\bibinfo {author} {\bibfnamefont {O.~H.~E.}\ \bibnamefont {{Philcox}}}\ and\ \bibinfo {author} {\bibfnamefont {J.}~\bibnamefont {{Ereza}}},\ }\href {https://doi.org/10.48550/arXiv.2401.09523} {\bibfield  {journal} {\bibinfo  {journal} {arXiv e-prints}\ ,\ \bibinfo {eid} {arXiv:2401.09523}} (\bibinfo {year} {2024})},\ \Eprint {https://arxiv.org/abs/2401.09523} {arXiv:2401.09523 [astro-ph.CO]} \BibitemShut {NoStop}%
\bibitem [{\citenamefont {{Adari}}\ and\ \citenamefont {{Slosar}}(2024)}]{2024arXiv240504660A}%
  \BibitemOpen
  \bibfield  {author} {\bibinfo {author} {\bibfnamefont {P.}~\bibnamefont {{Adari}}}\ and\ \bibinfo {author} {\bibfnamefont {A.}~\bibnamefont {{Slosar}}},\ }\href {https://doi.org/10.48550/arXiv.2405.04660} {\bibfield  {journal} {\bibinfo  {journal} {arXiv e-prints}\ ,\ \bibinfo {eid} {arXiv:2405.04660}} (\bibinfo {year} {2024})},\ \Eprint {https://arxiv.org/abs/2405.04660} {arXiv:2405.04660 [astro-ph.CO]} \BibitemShut {NoStop}%
\bibitem [{\citenamefont {{Krolewski}}\ \emph {et~al.}(2024)\citenamefont {{Krolewski}}, \citenamefont {{May}}, \citenamefont {{Smith}},\ and\ \citenamefont {{Hopkins}}}]{2024arXiv240703397K}%
  \BibitemOpen
  \bibfield  {author} {\bibinfo {author} {\bibfnamefont {A.}~\bibnamefont {{Krolewski}}}, \bibinfo {author} {\bibfnamefont {S.}~\bibnamefont {{May}}}, \bibinfo {author} {\bibfnamefont {K.}~\bibnamefont {{Smith}}},\ and\ \bibinfo {author} {\bibfnamefont {H.}~\bibnamefont {{Hopkins}}},\ }\href {https://doi.org/10.48550/arXiv.2407.03397} {\bibfield  {journal} {\bibinfo  {journal} {arXiv e-prints}\ ,\ \bibinfo {eid} {arXiv:2407.03397}} (\bibinfo {year} {2024})},\ \Eprint {https://arxiv.org/abs/2407.03397} {arXiv:2407.03397 [astro-ph.CO]} \BibitemShut {NoStop}%
\bibitem [{\citenamefont {{Philcox}}(2023)}]{2023PhRvL.131r1001P}%
  \BibitemOpen
  \bibfield  {author} {\bibinfo {author} {\bibfnamefont {O.~H.~E.}\ \bibnamefont {{Philcox}}},\ }\href {https://doi.org/10.1103/PhysRevLett.131.181001} {\bibfield  {journal} {\bibinfo  {journal} {\prl}\ }\textbf {\bibinfo {volume} {131}},\ \bibinfo {eid} {181001} (\bibinfo {year} {2023})},\ \Eprint {https://arxiv.org/abs/2303.12106} {arXiv:2303.12106 [astro-ph.CO]} \BibitemShut {NoStop}%
\bibitem [{\citenamefont {{Philcox}}\ and\ \citenamefont {{Shiraishi}}(2024)}]{2024PhRvD.109h3514P}%
  \BibitemOpen
  \bibfield  {author} {\bibinfo {author} {\bibfnamefont {O.~H.~E.}\ \bibnamefont {{Philcox}}}\ and\ \bibinfo {author} {\bibfnamefont {M.}~\bibnamefont {{Shiraishi}}},\ }\href {https://doi.org/10.1103/PhysRevD.109.083514} {\bibfield  {journal} {\bibinfo  {journal} {\prd}\ }\textbf {\bibinfo {volume} {109}},\ \bibinfo {eid} {083514} (\bibinfo {year} {2024})},\ \Eprint {https://arxiv.org/abs/2308.03831} {arXiv:2308.03831 [astro-ph.CO]} \BibitemShut {NoStop}%
\bibitem [{\citenamefont {{Motloch}}\ \emph {et~al.}(2022)\citenamefont {{Motloch}}, \citenamefont {{Pen}},\ and\ \citenamefont {{Yu}}}]{2022PhRvD.105h3512M}%
  \BibitemOpen
  \bibfield  {author} {\bibinfo {author} {\bibfnamefont {P.}~\bibnamefont {{Motloch}}}, \bibinfo {author} {\bibfnamefont {U.-L.}\ \bibnamefont {{Pen}}},\ and\ \bibinfo {author} {\bibfnamefont {H.-R.}\ \bibnamefont {{Yu}}},\ }\href {https://doi.org/10.1103/PhysRevD.105.083512} {\bibfield  {journal} {\bibinfo  {journal} {\prd}\ }\textbf {\bibinfo {volume} {105}},\ \bibinfo {eid} {083512} (\bibinfo {year} {2022})},\ \Eprint {https://arxiv.org/abs/2111.12590} {arXiv:2111.12590 [astro-ph.CO]} \BibitemShut {NoStop}%
\bibitem [{\citenamefont {{Minami}}\ and\ \citenamefont {{Komatsu}}(2020)}]{2020PhRvL.125v1301M}%
  \BibitemOpen
  \bibfield  {author} {\bibinfo {author} {\bibfnamefont {Y.}~\bibnamefont {{Minami}}}\ and\ \bibinfo {author} {\bibfnamefont {E.}~\bibnamefont {{Komatsu}}},\ }\href {https://doi.org/10.1103/PhysRevLett.125.221301} {\bibfield  {journal} {\bibinfo  {journal} {\prl}\ }\textbf {\bibinfo {volume} {125}},\ \bibinfo {eid} {221301} (\bibinfo {year} {2020})},\ \Eprint {https://arxiv.org/abs/2011.11254} {arXiv:2011.11254 [astro-ph.CO]} \BibitemShut {NoStop}%
\bibitem [{\citenamefont {{Diego-Palazuelos}}\ \emph {et~al.}(2022)\citenamefont {{Diego-Palazuelos}}, \citenamefont {{Eskilt}}, \citenamefont {{Minami}}, \citenamefont {{Tristram}}, \citenamefont {{Sullivan}}, \citenamefont {{Banday}}, \citenamefont {{Barreiro}}, \citenamefont {{Eriksen}}, \citenamefont {{G{\'o}rski}}, \citenamefont {{Keskitalo}}, \citenamefont {{Komatsu}}, \citenamefont {{Mart{\'\i}nez-Gonz{\'a}lez}}, \citenamefont {{Scott}}, \citenamefont {{Vielva}},\ and\ \citenamefont {{Wehus}}}]{2022PhRvL.128i1302D}%
  \BibitemOpen
  \bibfield  {author} {\bibinfo {author} {\bibfnamefont {P.}~\bibnamefont {{Diego-Palazuelos}}}, \bibinfo {author} {\bibfnamefont {J.~R.}\ \bibnamefont {{Eskilt}}}, \bibinfo {author} {\bibfnamefont {Y.}~\bibnamefont {{Minami}}}, \bibinfo {author} {\bibfnamefont {M.}~\bibnamefont {{Tristram}}}, \bibinfo {author} {\bibfnamefont {R.~M.}\ \bibnamefont {{Sullivan}}}, \bibinfo {author} {\bibfnamefont {A.~J.}\ \bibnamefont {{Banday}}}, \bibinfo {author} {\bibfnamefont {R.~B.}\ \bibnamefont {{Barreiro}}}, \bibinfo {author} {\bibfnamefont {H.~K.}\ \bibnamefont {{Eriksen}}}, \bibinfo {author} {\bibfnamefont {K.~M.}\ \bibnamefont {{G{\'o}rski}}}, \bibinfo {author} {\bibfnamefont {R.}~\bibnamefont {{Keskitalo}}}, \bibinfo {author} {\bibfnamefont {E.}~\bibnamefont {{Komatsu}}}, \bibinfo {author} {\bibfnamefont {E.}~\bibnamefont {{Mart{\'\i}nez-Gonz{\'a}lez}}}, \bibinfo {author} {\bibfnamefont {D.}~\bibnamefont {{Scott}}}, \bibinfo {author} {\bibfnamefont {P.}~\bibnamefont {{Vielva}}},\ and\ \bibinfo {author}
  {\bibfnamefont {I.~K.}\ \bibnamefont {{Wehus}}},\ }\href {https://doi.org/10.1103/PhysRevLett.128.091302} {\bibfield  {journal} {\bibinfo  {journal} {\prl}\ }\textbf {\bibinfo {volume} {128}},\ \bibinfo {eid} {091302} (\bibinfo {year} {2022})},\ \Eprint {https://arxiv.org/abs/2201.07682} {arXiv:2201.07682 [astro-ph.CO]} \BibitemShut {NoStop}%
\bibitem [{\citenamefont {{Eskilt}}\ and\ \citenamefont {{Komatsu}}(2022)}]{2022PhRvD.106f3503E}%
  \BibitemOpen
  \bibfield  {author} {\bibinfo {author} {\bibfnamefont {J.~R.}\ \bibnamefont {{Eskilt}}}\ and\ \bibinfo {author} {\bibfnamefont {E.}~\bibnamefont {{Komatsu}}},\ }\href {https://doi.org/10.1103/PhysRevD.106.063503} {\bibfield  {journal} {\bibinfo  {journal} {\prd}\ }\textbf {\bibinfo {volume} {106}},\ \bibinfo {eid} {063503} (\bibinfo {year} {2022})},\ \Eprint {https://arxiv.org/abs/2205.13962} {arXiv:2205.13962 [astro-ph.CO]} \BibitemShut {NoStop}%
\bibitem [{\citenamefont {{Masui}}\ and\ \citenamefont {{Pen}}(2010)}]{2010PhRvL.105p1302M}%
  \BibitemOpen
  \bibfield  {author} {\bibinfo {author} {\bibfnamefont {K.~W.}\ \bibnamefont {{Masui}}}\ and\ \bibinfo {author} {\bibfnamefont {U.-L.}\ \bibnamefont {{Pen}}},\ }\href {https://doi.org/10.1103/PhysRevLett.105.161302} {\bibfield  {journal} {\bibinfo  {journal} {\prl}\ }\textbf {\bibinfo {volume} {105}},\ \bibinfo {eid} {161302} (\bibinfo {year} {2010})},\ \Eprint {https://arxiv.org/abs/1006.4181} {arXiv:1006.4181 [astro-ph.CO]} \BibitemShut {NoStop}%
\bibitem [{\citenamefont {{Jeong}}\ and\ \citenamefont {{Kamionkowski}}(2012)}]{2012PhRvL.108y1301J}%
  \BibitemOpen
  \bibfield  {author} {\bibinfo {author} {\bibfnamefont {D.}~\bibnamefont {{Jeong}}}\ and\ \bibinfo {author} {\bibfnamefont {M.}~\bibnamefont {{Kamionkowski}}},\ }\href {https://doi.org/10.1103/PhysRevLett.108.251301} {\bibfield  {journal} {\bibinfo  {journal} {\prl}\ }\textbf {\bibinfo {volume} {108}},\ \bibinfo {eid} {251301} (\bibinfo {year} {2012})},\ \Eprint {https://arxiv.org/abs/1203.0302} {arXiv:1203.0302 [astro-ph.CO]} \BibitemShut {NoStop}%
\bibitem [{\citenamefont {{Masui}}\ \emph {et~al.}(2017)\citenamefont {{Masui}}, \citenamefont {{Pen}},\ and\ \citenamefont {{Turok}}}]{2017PhRvL.118v1301M}%
  \BibitemOpen
  \bibfield  {author} {\bibinfo {author} {\bibfnamefont {K.~W.}\ \bibnamefont {{Masui}}}, \bibinfo {author} {\bibfnamefont {U.-L.}\ \bibnamefont {{Pen}}},\ and\ \bibinfo {author} {\bibfnamefont {N.}~\bibnamefont {{Turok}}},\ }\href {https://doi.org/10.1103/PhysRevLett.118.221301} {\bibfield  {journal} {\bibinfo  {journal} {\prl}\ }\textbf {\bibinfo {volume} {118}},\ \bibinfo {eid} {221301} (\bibinfo {year} {2017})},\ \Eprint {https://arxiv.org/abs/1702.06552} {arXiv:1702.06552 [astro-ph.CO]} \BibitemShut {NoStop}%
\bibitem [{\citenamefont {{Coulton}}\ \emph {et~al.}(2024)\citenamefont {{Coulton}}, \citenamefont {{Philcox}},\ and\ \citenamefont {{Villaescusa-Navarro}}}]{2024PhRvD.109b3531C}%
  \BibitemOpen
  \bibfield  {author} {\bibinfo {author} {\bibfnamefont {W.~R.}\ \bibnamefont {{Coulton}}}, \bibinfo {author} {\bibfnamefont {O.~H.~E.}\ \bibnamefont {{Philcox}}},\ and\ \bibinfo {author} {\bibfnamefont {F.}~\bibnamefont {{Villaescusa-Navarro}}},\ }\href {https://doi.org/10.1103/PhysRevD.109.023531} {\bibfield  {journal} {\bibinfo  {journal} {\prd}\ }\textbf {\bibinfo {volume} {109}},\ \bibinfo {eid} {023531} (\bibinfo {year} {2024})},\ \Eprint {https://arxiv.org/abs/2306.11782} {arXiv:2306.11782 [astro-ph.CO]} \BibitemShut {NoStop}%
\bibitem [{\citenamefont {{Shim}}\ \emph {et~al.}(2024)\citenamefont {{Shim}}, \citenamefont {{Pen}}, \citenamefont {{Yu}},\ and\ \citenamefont {{Okumura}}}]{2024arXiv240606080S}%
  \BibitemOpen
  \bibfield  {author} {\bibinfo {author} {\bibfnamefont {J.}~\bibnamefont {{Shim}}}, \bibinfo {author} {\bibfnamefont {U.-L.}\ \bibnamefont {{Pen}}}, \bibinfo {author} {\bibfnamefont {H.-R.}\ \bibnamefont {{Yu}}},\ and\ \bibinfo {author} {\bibfnamefont {T.}~\bibnamefont {{Okumura}}},\ }\href {https://doi.org/10.48550/arXiv.2406.06080} {\bibfield  {journal} {\bibinfo  {journal} {arXiv e-prints}\ ,\ \bibinfo {eid} {arXiv:2406.06080}} (\bibinfo {year} {2024})},\ \Eprint {https://arxiv.org/abs/2406.06080} {arXiv:2406.06080 [astro-ph.CO]} \BibitemShut {NoStop}%
\bibitem [{\citenamefont {{Dai}}\ \emph {et~al.}(2013)\citenamefont {{Dai}}, \citenamefont {{Jeong}},\ and\ \citenamefont {{Kamionkowski}}}]{2013PhRvD..88d3507D}%
  \BibitemOpen
  \bibfield  {author} {\bibinfo {author} {\bibfnamefont {L.}~\bibnamefont {{Dai}}}, \bibinfo {author} {\bibfnamefont {D.}~\bibnamefont {{Jeong}}},\ and\ \bibinfo {author} {\bibfnamefont {M.}~\bibnamefont {{Kamionkowski}}},\ }\href {https://doi.org/10.1103/PhysRevD.88.043507} {\bibfield  {journal} {\bibinfo  {journal} {\prd}\ }\textbf {\bibinfo {volume} {88}},\ \bibinfo {eid} {043507} (\bibinfo {year} {2013})},\ \Eprint {https://arxiv.org/abs/1306.3985} {arXiv:1306.3985 [astro-ph.CO]} \BibitemShut {NoStop}%
\bibitem [{\citenamefont {{Schmidt}}\ \emph {et~al.}(2014)\citenamefont {{Schmidt}}, \citenamefont {{Pajer}},\ and\ \citenamefont {{Zaldarriaga}}}]{2014PhRvD..89h3507S}%
  \BibitemOpen
  \bibfield  {author} {\bibinfo {author} {\bibfnamefont {F.}~\bibnamefont {{Schmidt}}}, \bibinfo {author} {\bibfnamefont {E.}~\bibnamefont {{Pajer}}},\ and\ \bibinfo {author} {\bibfnamefont {M.}~\bibnamefont {{Zaldarriaga}}},\ }\href {https://doi.org/10.1103/PhysRevD.89.083507} {\bibfield  {journal} {\bibinfo  {journal} {\prd}\ }\textbf {\bibinfo {volume} {89}},\ \bibinfo {eid} {083507} (\bibinfo {year} {2014})},\ \Eprint {https://arxiv.org/abs/1312.5616} {arXiv:1312.5616 [astro-ph.CO]} \BibitemShut {NoStop}%
\bibitem [{\citenamefont {{Akitsu}}\ \emph {et~al.}(2023)\citenamefont {{Akitsu}}, \citenamefont {{Li}},\ and\ \citenamefont {{Okumura}}}]{2023PhRvD.107f3531A}%
  \BibitemOpen
  \bibfield  {author} {\bibinfo {author} {\bibfnamefont {K.}~\bibnamefont {{Akitsu}}}, \bibinfo {author} {\bibfnamefont {Y.}~\bibnamefont {{Li}}},\ and\ \bibinfo {author} {\bibfnamefont {T.}~\bibnamefont {{Okumura}}},\ }\href {https://doi.org/10.1103/PhysRevD.107.063531} {\bibfield  {journal} {\bibinfo  {journal} {\prd}\ }\textbf {\bibinfo {volume} {107}},\ \bibinfo {eid} {063531} (\bibinfo {year} {2023})},\ \Eprint {https://arxiv.org/abs/2209.06226} {arXiv:2209.06226 [astro-ph.CO]} \BibitemShut {NoStop}%
\bibitem [{\citenamefont {{Caldwell}}\ and\ \citenamefont {{Devulder}}(2018)}]{2018PhRvD..97b3532C}%
  \BibitemOpen
  \bibfield  {author} {\bibinfo {author} {\bibfnamefont {R.~R.}\ \bibnamefont {{Caldwell}}}\ and\ \bibinfo {author} {\bibfnamefont {C.}~\bibnamefont {{Devulder}}},\ }\href {https://doi.org/10.1103/PhysRevD.97.023532} {\bibfield  {journal} {\bibinfo  {journal} {\prd}\ }\textbf {\bibinfo {volume} {97}},\ \bibinfo {eid} {023532} (\bibinfo {year} {2018})},\ \Eprint {https://arxiv.org/abs/1706.03765} {arXiv:1706.03765 [astro-ph.CO]} \BibitemShut {NoStop}%
\bibitem [{\citenamefont {{Zhu}}\ and\ \citenamefont {{Pen}}(2020)}]{2020arXiv201108251Z}%
  \BibitemOpen
  \bibfield  {author} {\bibinfo {author} {\bibfnamefont {H.-M.}\ \bibnamefont {{Zhu}}}\ and\ \bibinfo {author} {\bibfnamefont {U.-L.}\ \bibnamefont {{Pen}}},\ }\href {https://doi.org/10.48550/arXiv.2011.08251} {\bibfield  {journal} {\bibinfo  {journal} {arXiv e-prints}\ ,\ \bibinfo {eid} {arXiv:2011.08251}} (\bibinfo {year} {2020})},\ \Eprint {https://arxiv.org/abs/2011.08251} {arXiv:2011.08251 [astro-ph.CO]} \BibitemShut {NoStop}%
\bibitem [{\citenamefont {{Pen}}(1997)}]{1997ApJ...490L.127P}%
  \BibitemOpen
  \bibfield  {author} {\bibinfo {author} {\bibfnamefont {U.-L.}\ \bibnamefont {{Pen}}},\ }\href {https://doi.org/10.1086/311042} {\bibfield  {journal} {\bibinfo  {journal} {\apjl}\ }\textbf {\bibinfo {volume} {490}},\ \bibinfo {pages} {L127} (\bibinfo {year} {1997})},\ \Eprint {https://arxiv.org/abs/astro-ph/9709261} {arXiv:astro-ph/9709261 [astro-ph]} \BibitemShut {NoStop}%
\bibitem [{\citenamefont {{Trac}}\ and\ \citenamefont {{Pen}}(2006)}]{2006NewA...11..273T}%
  \BibitemOpen
  \bibfield  {author} {\bibinfo {author} {\bibfnamefont {H.}~\bibnamefont {{Trac}}}\ and\ \bibinfo {author} {\bibfnamefont {U.-L.}\ \bibnamefont {{Pen}}},\ }\href {https://doi.org/10.1016/j.newast.2005.08.003} {\bibfield  {journal} {\bibinfo  {journal} {\na}\ }\textbf {\bibinfo {volume} {11}},\ \bibinfo {pages} {273} (\bibinfo {year} {2006})},\ \Eprint {https://arxiv.org/abs/astro-ph/0402444} {arXiv:astro-ph/0402444 [astro-ph]} \BibitemShut {NoStop}%
\bibitem [{\citenamefont {{Harnois-D{\'e}raps}}\ \emph {et~al.}(2013)\citenamefont {{Harnois-D{\'e}raps}}, \citenamefont {{Pen}}, \citenamefont {{Iliev}}, \citenamefont {{Merz}}, \citenamefont {{Emberson}},\ and\ \citenamefont {{Desjacques}}}]{2013MNRAS.436..540H}%
  \BibitemOpen
  \bibfield  {author} {\bibinfo {author} {\bibfnamefont {J.}~\bibnamefont {{Harnois-D{\'e}raps}}}, \bibinfo {author} {\bibfnamefont {U.-L.}\ \bibnamefont {{Pen}}}, \bibinfo {author} {\bibfnamefont {I.~T.}\ \bibnamefont {{Iliev}}}, \bibinfo {author} {\bibfnamefont {H.}~\bibnamefont {{Merz}}}, \bibinfo {author} {\bibfnamefont {J.~D.}\ \bibnamefont {{Emberson}}},\ and\ \bibinfo {author} {\bibfnamefont {V.}~\bibnamefont {{Desjacques}}},\ }\href {https://doi.org/10.1093/mnras/stt1591} {\bibfield  {journal} {\bibinfo  {journal} {\mnras}\ }\textbf {\bibinfo {volume} {436}},\ \bibinfo {pages} {540} (\bibinfo {year} {2013})},\ \Eprint {https://arxiv.org/abs/1208.5098} {arXiv:1208.5098 [astro-ph.CO]} \BibitemShut {NoStop}%
\bibitem [{\citenamefont {{Pen}}\ \emph {et~al.}(2012)\citenamefont {{Pen}}, \citenamefont {{Sheth}}, \citenamefont {{Harnois-Deraps}}, \citenamefont {{Chen}},\ and\ \citenamefont {{Li}}}]{2012arXiv1202.5804P}%
  \BibitemOpen
  \bibfield  {author} {\bibinfo {author} {\bibfnamefont {U.-L.}\ \bibnamefont {{Pen}}}, \bibinfo {author} {\bibfnamefont {R.}~\bibnamefont {{Sheth}}}, \bibinfo {author} {\bibfnamefont {J.}~\bibnamefont {{Harnois-Deraps}}}, \bibinfo {author} {\bibfnamefont {X.}~\bibnamefont {{Chen}}},\ and\ \bibinfo {author} {\bibfnamefont {Z.}~\bibnamefont {{Li}}},\ }\href {https://doi.org/10.48550/arXiv.1202.5804} {\bibfield  {journal} {\bibinfo  {journal} {arXiv e-prints}\ ,\ \bibinfo {eid} {arXiv:1202.5804}} (\bibinfo {year} {2012})},\ \Eprint {https://arxiv.org/abs/1202.5804} {arXiv:1202.5804 [astro-ph.CO]} \BibitemShut {NoStop}%
\bibitem [{\citenamefont {{Zhu}}\ \emph {et~al.}(2016)\citenamefont {{Zhu}}, \citenamefont {{Pen}}, \citenamefont {{Yu}}, \citenamefont {{Er}},\ and\ \citenamefont {{Chen}}}]{2016PhRvD..93j3504Z}%
  \BibitemOpen
  \bibfield  {author} {\bibinfo {author} {\bibfnamefont {H.-M.}\ \bibnamefont {{Zhu}}}, \bibinfo {author} {\bibfnamefont {U.-L.}\ \bibnamefont {{Pen}}}, \bibinfo {author} {\bibfnamefont {Y.}~\bibnamefont {{Yu}}}, \bibinfo {author} {\bibfnamefont {X.}~\bibnamefont {{Er}}},\ and\ \bibinfo {author} {\bibfnamefont {X.}~\bibnamefont {{Chen}}},\ }\href {https://doi.org/10.1103/PhysRevD.93.103504} {\bibfield  {journal} {\bibinfo  {journal} {\prd}\ }\textbf {\bibinfo {volume} {93}},\ \bibinfo {eid} {103504} (\bibinfo {year} {2016})},\ \Eprint {https://arxiv.org/abs/1511.04680} {arXiv:1511.04680 [astro-ph.CO]} \BibitemShut {NoStop}%
\bibitem [{\citenamefont {{Kara{\c{c}}ayl{\i}}}\ and\ \citenamefont {{Padmanabhan}}(2019)}]{2019MNRAS.486.3864K}%
  \BibitemOpen
  \bibfield  {author} {\bibinfo {author} {\bibfnamefont {N.~G.}\ \bibnamefont {{Kara{\c{c}}ayl{\i}}}}\ and\ \bibinfo {author} {\bibfnamefont {N.}~\bibnamefont {{Padmanabhan}}},\ }\href {https://doi.org/10.1093/mnras/stz964} {\bibfield  {journal} {\bibinfo  {journal} {\mnras}\ }\textbf {\bibinfo {volume} {486}},\ \bibinfo {pages} {3864} (\bibinfo {year} {2019})},\ \Eprint {https://arxiv.org/abs/1904.01387} {arXiv:1904.01387 [astro-ph.CO]} \BibitemShut {NoStop}%
\bibitem [{\citenamefont {{Zhu}}\ \emph {et~al.}(2022)\citenamefont {{Zhu}}, \citenamefont {{Mao}},\ and\ \citenamefont {{Pen}}}]{2022ApJ...929....5Z}%
  \BibitemOpen
  \bibfield  {author} {\bibinfo {author} {\bibfnamefont {H.-M.}\ \bibnamefont {{Zhu}}}, \bibinfo {author} {\bibfnamefont {T.-X.}\ \bibnamefont {{Mao}}},\ and\ \bibinfo {author} {\bibfnamefont {U.-L.}\ \bibnamefont {{Pen}}},\ }\href {https://doi.org/10.3847/1538-4357/ac5a47} {\bibfield  {journal} {\bibinfo  {journal} {\apj}\ }\textbf {\bibinfo {volume} {929}},\ \bibinfo {eid} {5} (\bibinfo {year} {2022})},\ \Eprint {https://arxiv.org/abs/2108.01575} {arXiv:2108.01575 [astro-ph.CO]} \BibitemShut {NoStop}%
\bibitem [{\citenamefont {{Zang}}\ \emph {et~al.}(2024)\citenamefont {{Zang}}, \citenamefont {{Zhu}}, \citenamefont {{Schmittfull}},\ and\ \citenamefont {{Pen}}}]{2024ApJ...962...21Z}%
  \BibitemOpen
  \bibfield  {author} {\bibinfo {author} {\bibfnamefont {S.-H.}\ \bibnamefont {{Zang}}}, \bibinfo {author} {\bibfnamefont {H.-M.}\ \bibnamefont {{Zhu}}}, \bibinfo {author} {\bibfnamefont {M.}~\bibnamefont {{Schmittfull}}},\ and\ \bibinfo {author} {\bibfnamefont {U.-L.}\ \bibnamefont {{Pen}}},\ }\href {https://doi.org/10.3847/1538-4357/ad0cf0} {\bibfield  {journal} {\bibinfo  {journal} {\apj}\ }\textbf {\bibinfo {volume} {962}},\ \bibinfo {eid} {21} (\bibinfo {year} {2024})},\ \Eprint {https://arxiv.org/abs/2212.04294} {arXiv:2212.04294 [astro-ph.CO]} \BibitemShut {NoStop}%
\bibitem [{\citenamefont {{Foreman}}\ \emph {et~al.}(2018)\citenamefont {{Foreman}}, \citenamefont {{Meerburg}}, \citenamefont {{van Engelen}},\ and\ \citenamefont {{Meyers}}}]{2018JCAP...07..046F}%
  \BibitemOpen
  \bibfield  {author} {\bibinfo {author} {\bibfnamefont {S.}~\bibnamefont {{Foreman}}}, \bibinfo {author} {\bibfnamefont {P.~D.}\ \bibnamefont {{Meerburg}}}, \bibinfo {author} {\bibfnamefont {A.}~\bibnamefont {{van Engelen}}},\ and\ \bibinfo {author} {\bibfnamefont {J.}~\bibnamefont {{Meyers}}},\ }\href {https://doi.org/10.1088/1475-7516/2018/07/046} {\bibfield  {journal} {\bibinfo  {journal} {\jcap}\ }\textbf {\bibinfo {volume} {2018}},\ \bibinfo {eid} {046} (\bibinfo {year} {2018})},\ \Eprint {https://arxiv.org/abs/1803.04975} {arXiv:1803.04975 [astro-ph.CO]} \BibitemShut {NoStop}%
\bibitem [{\citenamefont {{Li}}\ \emph {et~al.}(2020{\natexlab{a}})\citenamefont {{Li}}, \citenamefont {{Dodelson}},\ and\ \citenamefont {{Croft}}}]{2020PhRvD.101h3510L}%
  \BibitemOpen
  \bibfield  {author} {\bibinfo {author} {\bibfnamefont {P.}~\bibnamefont {{Li}}}, \bibinfo {author} {\bibfnamefont {S.}~\bibnamefont {{Dodelson}}},\ and\ \bibinfo {author} {\bibfnamefont {R.~A.~C.}\ \bibnamefont {{Croft}}},\ }\href {https://doi.org/10.1103/PhysRevD.101.083510} {\bibfield  {journal} {\bibinfo  {journal} {\prd}\ }\textbf {\bibinfo {volume} {101}},\ \bibinfo {eid} {083510} (\bibinfo {year} {2020}{\natexlab{a}})},\ \Eprint {https://arxiv.org/abs/2001.02780} {arXiv:2001.02780 [astro-ph.CO]} \BibitemShut {NoStop}%
\bibitem [{\citenamefont {{Li}}\ \emph {et~al.}(2020{\natexlab{b}})\citenamefont {{Li}}, \citenamefont {{Croft}},\ and\ \citenamefont {{Dodelson}}}]{2020arXiv200700226L}%
  \BibitemOpen
  \bibfield  {author} {\bibinfo {author} {\bibfnamefont {P.}~\bibnamefont {{Li}}}, \bibinfo {author} {\bibfnamefont {R.~A.~C.}\ \bibnamefont {{Croft}}},\ and\ \bibinfo {author} {\bibfnamefont {S.}~\bibnamefont {{Dodelson}}},\ }\href {https://doi.org/10.48550/arXiv.2007.00226} {\bibfield  {journal} {\bibinfo  {journal} {arXiv e-prints}\ ,\ \bibinfo {eid} {arXiv:2007.00226}} (\bibinfo {year} {2020}{\natexlab{b}})},\ \Eprint {https://arxiv.org/abs/2007.00226} {arXiv:2007.00226 [astro-ph.CO]} \BibitemShut {NoStop}%
\bibitem [{\citenamefont {{Darwish}}\ \emph {et~al.}(2021)\citenamefont {{Darwish}}, \citenamefont {{Foreman}}, \citenamefont {{Abidi}}, \citenamefont {{Baldauf}}, \citenamefont {{Sherwin}},\ and\ \citenamefont {{Meerburg}}}]{2021PhRvD.104l3520D}%
  \BibitemOpen
  \bibfield  {author} {\bibinfo {author} {\bibfnamefont {O.}~\bibnamefont {{Darwish}}}, \bibinfo {author} {\bibfnamefont {S.}~\bibnamefont {{Foreman}}}, \bibinfo {author} {\bibfnamefont {M.~M.}\ \bibnamefont {{Abidi}}}, \bibinfo {author} {\bibfnamefont {T.}~\bibnamefont {{Baldauf}}}, \bibinfo {author} {\bibfnamefont {B.~D.}\ \bibnamefont {{Sherwin}}},\ and\ \bibinfo {author} {\bibfnamefont {P.~D.}\ \bibnamefont {{Meerburg}}},\ }\href {https://doi.org/10.1103/PhysRevD.104.123520} {\bibfield  {journal} {\bibinfo  {journal} {\prd}\ }\textbf {\bibinfo {volume} {104}},\ \bibinfo {eid} {123520} (\bibinfo {year} {2021})},\ \Eprint {https://arxiv.org/abs/2007.08472} {arXiv:2007.08472 [astro-ph.CO]} \BibitemShut {NoStop}%
\bibitem [{\citenamefont {{Wang}}\ and\ \citenamefont {{Jeong}}(2024)}]{2024JCAP...07..020W}%
  \BibitemOpen
  \bibfield  {author} {\bibinfo {author} {\bibfnamefont {Z.}~\bibnamefont {{Wang}}}\ and\ \bibinfo {author} {\bibfnamefont {D.}~\bibnamefont {{Jeong}}},\ }\href {https://doi.org/10.1088/1475-7516/2024/07/020} {\bibfield  {journal} {\bibinfo  {journal} {\jcap}\ }\textbf {\bibinfo {volume} {2024}},\ \bibinfo {eid} {020} (\bibinfo {year} {2024})},\ \Eprint {https://arxiv.org/abs/2312.17321} {arXiv:2312.17321 [astro-ph.CO]} \BibitemShut {NoStop}%
\bibitem [{\citenamefont {Cahn}\ \emph {et~al.}(2023)\citenamefont {Cahn}, \citenamefont {Slepian},\ and\ \citenamefont {Hou}}]{Cahn:2021ltp}%
  \BibitemOpen
  \bibfield  {author} {\bibinfo {author} {\bibfnamefont {R.~N.}\ \bibnamefont {Cahn}}, \bibinfo {author} {\bibfnamefont {Z.}~\bibnamefont {Slepian}},\ and\ \bibinfo {author} {\bibfnamefont {J.}~\bibnamefont {Hou}},\ }\href {https://doi.org/10.1103/PhysRevLett.130.201002} {\bibfield  {journal} {\bibinfo  {journal} {Phys. Rev. Lett.}\ }\textbf {\bibinfo {volume} {130}},\ \bibinfo {pages} {201002} (\bibinfo {year} {2023})},\ \Eprint {https://arxiv.org/abs/2110.12004} {arXiv:2110.12004 [astro-ph.CO]} \BibitemShut {NoStop}%
\bibitem [{\citenamefont {{Yu}}\ \emph {et~al.}(2020)\citenamefont {{Yu}}, \citenamefont {{Motloch}}, \citenamefont {{Pen}}, \citenamefont {{Yu}}, \citenamefont {{Wang}}, \citenamefont {{Mo}}, \citenamefont {{Yang}},\ and\ \citenamefont {{Jing}}}]{2020PhRvL.124j1302Y}%
  \BibitemOpen
  \bibfield  {author} {\bibinfo {author} {\bibfnamefont {H.-R.}\ \bibnamefont {{Yu}}}, \bibinfo {author} {\bibfnamefont {P.}~\bibnamefont {{Motloch}}}, \bibinfo {author} {\bibfnamefont {U.-L.}\ \bibnamefont {{Pen}}}, \bibinfo {author} {\bibfnamefont {Y.}~\bibnamefont {{Yu}}}, \bibinfo {author} {\bibfnamefont {H.}~\bibnamefont {{Wang}}}, \bibinfo {author} {\bibfnamefont {H.}~\bibnamefont {{Mo}}}, \bibinfo {author} {\bibfnamefont {X.}~\bibnamefont {{Yang}}},\ and\ \bibinfo {author} {\bibfnamefont {Y.}~\bibnamefont {{Jing}}},\ }\href {https://doi.org/10.1103/PhysRevLett.124.101302} {\bibfield  {journal} {\bibinfo  {journal} {\prl}\ }\textbf {\bibinfo {volume} {124}},\ \bibinfo {eid} {101302} (\bibinfo {year} {2020})},\ \Eprint {https://arxiv.org/abs/1904.01029} {arXiv:1904.01029 [astro-ph.CO]} \BibitemShut {NoStop}%
\bibitem [{\citenamefont {{Okumura}}\ and\ \citenamefont {{Sasaki}}(2024)}]{2024arXiv240504210O}%
  \BibitemOpen
  \bibfield  {author} {\bibinfo {author} {\bibfnamefont {T.}~\bibnamefont {{Okumura}}}\ and\ \bibinfo {author} {\bibfnamefont {M.}~\bibnamefont {{Sasaki}}},\ }\href {https://doi.org/10.48550/arXiv.2405.04210} {\bibfield  {journal} {\bibinfo  {journal} {arXiv e-prints}\ ,\ \bibinfo {eid} {arXiv:2405.04210}} (\bibinfo {year} {2024})},\ \Eprint {https://arxiv.org/abs/2405.04210} {arXiv:2405.04210 [astro-ph.CO]} \BibitemShut {NoStop}%
\bibitem [{\citenamefont {{Jamieson}}\ \emph {et~al.}(2024)\citenamefont {{Jamieson}}, \citenamefont {{Caravano}}, \citenamefont {{Hou}}, \citenamefont {{Slepian}},\ and\ \citenamefont {{Komatsu}}}]{2024arXiv240615683J}%
  \BibitemOpen
  \bibfield  {author} {\bibinfo {author} {\bibfnamefont {D.}~\bibnamefont {{Jamieson}}}, \bibinfo {author} {\bibfnamefont {A.}~\bibnamefont {{Caravano}}}, \bibinfo {author} {\bibfnamefont {J.}~\bibnamefont {{Hou}}}, \bibinfo {author} {\bibfnamefont {Z.}~\bibnamefont {{Slepian}}},\ and\ \bibinfo {author} {\bibfnamefont {E.}~\bibnamefont {{Komatsu}}},\ }\href {https://doi.org/10.48550/arXiv.2406.15683} {\bibfield  {journal} {\bibinfo  {journal} {arXiv e-prints}\ ,\ \bibinfo {eid} {arXiv:2406.15683}} (\bibinfo {year} {2024})},\ \Eprint {https://arxiv.org/abs/2406.15683} {arXiv:2406.15683 [astro-ph.CO]} \BibitemShut {NoStop}%
\bibitem [{\citenamefont {{Taylor}}\ \emph {et~al.}(2024)\citenamefont {{Taylor}}, \citenamefont {{Craigie}},\ and\ \citenamefont {{Ting}}}]{2024PhRvD.109h3518T}%
  \BibitemOpen
  \bibfield  {author} {\bibinfo {author} {\bibfnamefont {P.~L.}\ \bibnamefont {{Taylor}}}, \bibinfo {author} {\bibfnamefont {M.}~\bibnamefont {{Craigie}}},\ and\ \bibinfo {author} {\bibfnamefont {Y.-S.}\ \bibnamefont {{Ting}}},\ }\href {https://doi.org/10.1103/PhysRevD.109.083518} {\bibfield  {journal} {\bibinfo  {journal} {\prd}\ }\textbf {\bibinfo {volume} {109}},\ \bibinfo {eid} {083518} (\bibinfo {year} {2024})},\ \Eprint {https://arxiv.org/abs/2312.09287} {arXiv:2312.09287 [astro-ph.CO]} \BibitemShut {NoStop}%
\bibitem [{\citenamefont {{Craigie}}\ \emph {et~al.}(2024)\citenamefont {{Craigie}}, \citenamefont {{Taylor}}, \citenamefont {{Ting}}, \citenamefont {{Cuesta-Lazaro}}, \citenamefont {{Ruggeri}},\ and\ \citenamefont {{Davis}}}]{2024arXiv240513083C}%
  \BibitemOpen
  \bibfield  {author} {\bibinfo {author} {\bibfnamefont {M.}~\bibnamefont {{Craigie}}}, \bibinfo {author} {\bibfnamefont {P.~L.}\ \bibnamefont {{Taylor}}}, \bibinfo {author} {\bibfnamefont {Y.-S.}\ \bibnamefont {{Ting}}}, \bibinfo {author} {\bibfnamefont {C.}~\bibnamefont {{Cuesta-Lazaro}}}, \bibinfo {author} {\bibfnamefont {R.}~\bibnamefont {{Ruggeri}}},\ and\ \bibinfo {author} {\bibfnamefont {T.~M.}\ \bibnamefont {{Davis}}},\ }\href {https://doi.org/10.48550/arXiv.2405.13083} {\bibfield  {journal} {\bibinfo  {journal} {arXiv e-prints}\ ,\ \bibinfo {eid} {arXiv:2405.13083}} (\bibinfo {year} {2024})},\ \Eprint {https://arxiv.org/abs/2405.13083} {arXiv:2405.13083 [astro-ph.IM]} \BibitemShut {NoStop}%
\end{thebibliography}%

\bibliographystyle{apsrev4-2}

\end{document}